\documentstyle[preprint,tighten,aps,prb]{revtex}
\begin{document}
\def\LCO{La$_2$CuO$_4$}
\def\S2CO{Sr$_2$CuO$_3$}
\def\BICO{Bi$_2$CuO$_4$}
\def\LICO{Li$_2$CuO$_2$}
\def\SCO{SrCuO$_2$}
\def\BCOCL{Ba$_2$Cu$_3$O$_4$Cl$_2$}
\def\SCOCL{Sr$_2$CuO$_2$Cl$_2$}
\def\CASTEL{Ba$_3$Cu$_2$O$_4$Cl$_2$}
\def\UL{\underline L}
\def\D{$2p^5 3d^9$}  \def\L{$2p^5 3d^{10}$\UL}
\def\ZR{$2p^5 3d^{10}$}
\def\CUO{CuO$_2$}
\def\DEL{$\Delta$}
\def\UDC{$U_{dc}$}
\def\UDD{$U_{ dd}$}
\def\pds{$pd\sigma$}
\def\pps{$pp\sigma$}
\def\ppp{$pp\pi$}
\def\ISIM{$I_{s} / I_{m}$}
%
\draft
%
\title{Cu-O network dependent core hole screening in low-dimensional
cuprate systems: a high-resolution x-ray photoemission study}
\author{T. B\"oske, K. Maiti\cite{add1},
O. Knauff, K. Ruck,
M. S. Golden, G. Krabbes, and J. Fink}
\address{Institut f\"ur Festk\"orper- und Werkstofforschung
Dresden, Postfach 270016, 01171 Dresden, Germany}
\author{T. Osafune, N. Motoyama, H. Eisaki, and S. Uchida}
\address{Department of Superconductivity,  Faculty of Engineering,
University of Tokyo, Yayoi 2-11-16, Bunkyo-ku, Tokyo 113, Japan}
\maketitle
\begin{abstract}
We present an experimental study of the dynamics of
holes in the valence bands of 0, 1, and 2 dimensional
undoped model cuprates, as expressed via the screening
of a Cu $2p$ core hole.
The response depends strongly upon the dimensionality and the
details of the Cu-O-Cu network geometry and clearly goes
beyond the present theoretical state-of-the-art description
within the three-band $d$-$p$ model.
\end{abstract}
%
%
%
The dynamics of single hole states in low dimensional,
highly correlated electron systems are currently of
great interest in the context of the high-$T_C$ superconducting cuprates.
Recently, the band dispersion
of a single hole in an antiferromagnetic insulator
consisting of a two-dimensional (2D) Cu-O network
has been studied\cite{wells}
and has promoted  further investigations
in this direction.\cite{potje,goldie}
As regards  one-dimensional (1D) \CUO\ chain systems,
a separation between spin and charge excitations  most
unlike the planar \CUO\ systems
has recently been reported in the zigzag
chain cuprate \SCO.\cite{kim}

The application of core level
x-ray photoemission spectroscopy (XPS) to study
the electronic structure of correlated systems
is widely established, and from
an analysis of their Cu $2p$ core level line shapes
important contributions to
the understanding of the electronic structure of
the high-$T_C$ cuprates have been derived.\cite{kotani}
Experimentally, the Cu $2p$ photoelectron spectra of the formally
divalent copper compounds
show a so-called satellite emission at higher
binding energy than a
broad asymmetric main line. Such spectra
can be approximately described
with cluster calculations\cite{vdl} or
in an Anderson impurity model\cite{kotani,gunni}
using only one Cu site.
These single-site models discriminate between poorly screened
multiplets of the \D\ final state, mainly responsible
for the satellite and a
well screened \L\ final state for the
main line emission, where \UL\ denotes
that the intrinsic valence band hole is
situated in the ligands surrounding the Cu $2p$ core hole site.
The energy separation and the intensity ratio of these
spectral features are determined
by the valence band-core hole interaction energy
\UDC, the  effective Cu $3d$-O $2p$ hopping energy $t$, and
the charge transfer energy between a Cu $3d$ and an
O $2p$ state \DEL.\cite{vdl}
However, the experimentally observed large width and asymmetry
of the Cu $2p$ main line could not be understood within
these models.

To tackle this problem, the single-site models have been extended
to include different sites within the Cu-O network
by considering larger clusters.\cite{veenendaal} For a
linear Cu$_3$O$_{10}$ cluster, the Cu $2p$ main line was predicted
to consist of two components:\cite{veenendaal} the lowest binding
energy feature would be due to a so-called nonlocal screening
process where the valence band hole delocalizes
to form a Zhang-Rice singlet
(ZRS) state on CuO$_4$ plaquettes other than where the
core hole resides. The higher binding
energy component would then be a final state where the valence
band hole is predominantly located in the O $2p$
states immediately surrounding the core hole site and is consequently
called a locally screened final state.
In order to calculate the Cu $2p$ XPS of
larger systems (such as infinite chains or planes) further
approximations have to be made.\cite{okada95}
Such calculations appear to show, however,
that in general there remain considerable cluster-size
dependencies in the results. For example, the prominent double-featured
main line predicted for the linear Cu$_3$O$_{10}$ cluster\cite{veenendaal}
appears to be absent in larger 1D and 2D systems.\cite{okada95}
This is in agreement with 
previous Cu $2p$ photoemission data of different cuprates where no double
feature in the main line has been reported,\cite{tranquada,kalo97}
but it disagrees with a recent study of \SCOCL\ where the Cu $2p$
main line is shown to consist of at least three distinct
features.\cite{myself}

Therefore, a complete study of the Cu $2p$
photoemission as a function of both the
dimensionality and the Cu-O-Cu interaction geometry 
is needed
to put the multiple-site models on a sound basis and
to directly verify the different screening
channels experimentally.
Furthermore, based upon the multiple-site screening
approach, XPS
offers an elegant way to study
the dynamics of hole states in different Cu-O networks. 
In this paper we present the first systematic
study of Cu $2p$ photoemission in various
undoped cuprates with 0, 1, and 2D Cu-O networks.
In particular, using high-resolution XPS,
we examine high-quality single-crystals of \BICO, \LICO,
\CASTEL,  \S2CO, \SCO, \SCOCL, and \BCOCL.
These systems together represent a wide
variety of Cu-O network structures as depicted in
Fig.~\ref{network}.

For the 1D systems, we study
the extreme Cu-O-Cu interaction angles,
ranging from 180$^\circ$ in the linear chain \S2CO\ to nearly 90$^\circ$
in \LICO.
Both of these limiting Cu-O-Cu configurations play a
defining role in the physics of the spin-ladder systems\cite{dagotto} which
are currently of great interest.
We also include \BICO\ in this study as a representative
of a so-called zero-dimensional (0D) system. In this compound,
the CuO$_4$ plaquettes are
stacked along the $c$ direction with O-Bi-O
bridging units\cite{ong90} and no coherent Cu-O network
is realized.
\CASTEL\ contains chains of edge-shared CuO$_4$ plaquettes, which are
so arranged as to give a crenellated Cu-O network as
depicted in Fig.~\ref{network}.\cite{kipka76}

\SCOCL\ is considered to be the paradigm 
2D spin-$\frac{1}{2}$ Heisenberg antiferromagnet.\cite{scoclmagnet}
\BCOCL\ has a similar structure to that of \SCOCL,\cite{krabbes}
although its Cu-O network is composed of a regular
\CUO\ plane (denoted by Cu$_A$
in Fig.~\ref{network}) with an additional Cu site (Cu$_B$)
which is connected to the regular Cu$_A$O$_2$ plane
via a 90$^\circ$ Cu$_A$-O-Cu$_B$ configuration.
Consequently, the Cu atoms in \BCOCL\ show  two antiferromagnetic
phase-transitions at widely different temperatures
and two different ZRS dispersion functions
connected with these different Cu sub-systems.\cite{goldie}

The experiments were carried out
using a Perkin-Elmer photoemission system equipped
with a monochromatic Al K$\alpha$ source giving a resolution of 
about $0.4$ eV. Since the samples are insulating,
corrections for charging effects were undertaken
resulting in an estimated accuracy of the given absolute
energy values of $\pm 0.3$ eV.
The measurements were performed at room temperature
and the crystals were cleaved {\it in situ}
under ultrahigh vacuum conditions.
The O $1s$ spectra of all samples show negligible
emission at binding energies greater than $531$ eV,
which indicates clean samples.
We found that the shape of the Cu $2p$
spectra depends critically on the
cleanliness of the oxygen signal.

In Fig.~\ref{cu2p}  we show the
Cu $2p_\frac{3}{2}$ spectra of the cuprates.
An integral background has been subtracted and
the spectra are normalized to the leading
peak. The area of the \BCOCL\ 
spectrum is multiplied
by a factor of $1.5$ compared to that of
\SCOCL, assuming it to be
proportional to the number of Cu atoms in the
Cu-O plane for reasons to be shown later.
The spectral features for all crystals studied
are summarized in Table \ref{tab}.

In the following, we will
describe the salient features of the Cu $2p$ main lines
and then discuss the role of the electronic states
near the chemical potential in the screening processes responsible
for the observed structures.
In contrast to all
previous XPS studies known to us,
the Cu $2p$ main line spectra
show either a rich fine structure or are fairly
narrow symmetric lines.
The  position of the lowest
binding energy feature denoted by $A$ is the same
within our experimental accuracy for all the systems studied.
The spectrum of \BICO\ is very similar
to a previously published one.\cite{goldoni94}
A recent low resolution XPS study also gave a comparable spectrum for
polycrystalline \S2CO,\cite{kalo97} although, in the present case
using high resolution XPS, a 
shoulder denoted $C^\prime$ accompanying
the leading main line feature $A$
is clearly resolved.
The spectrum for the zigzag chain \SCO\ 
closely resembles that of \S2CO, whereby
the intensity of $C^\prime$ is larger in the former.
In addition, referring
to Table \ref{tab}, the satellite to main line
intensity ratio, \ISIM, is larger for \SCO\
than for \S2CO, indicating a larger \DEL\
for the zigzag chain in the language of single-site
models.
In contrast, for \BICO, \LICO, and \CASTEL\ a comparatively
narrow and symmetric main line $B$ is observed
at lower binding energy than
feature $C^\prime$ in the other 1D cuprates.
At the position of feature $A$ either no or small
spectral intensity is observed.

The spectrum
of \SCOCL\ has been discussed
in more detail elsewhere,\cite{myself}
The broad main line is
composed of two features
denoted  $A$ and $B$, and a prominent
shoulder $C$ appears at higher binding energies.
Compared to \SCOCL, the feature $B$ in \BCOCL\
is much more pronounced. For both 2D systems,
\ISIM\ is nearly the same whereas 
$C$  is positioned at different binding energies.
Significantly, the position of feature $B$ is very close
to that of the main lines of \BICO, \LICO, and \CASTEL.

To understand these features we will focus first
on the 0D and 1D systems and relate the main line
features to the differently coupled CuO$_4$ units
in \BICO, \LICO, \CASTEL, \SCO, and \S2CO, 
assuming the simplified geometry
of the Cu-O networks displayed in Fig.~\ref{network}.
In the case of \BICO, the interpretation
of the main line spectrum is straightforward:
since no coherent planar or linear Cu-O network exists,
only a local screening process is possible.
Similar arguments can be applied to \LICO\ as the
edge-sharing configuration of the CuO$_4$ plaquettes
with 90$^\circ$ Cu-O-Cu interactions does not support
direct oxygen-mediated hole hopping. The hopping processes are 
instead determined by
the small parameter $t_{pp}$.
The similarity of the spectrum of \CASTEL\ to that of \LICO\ shows that, 
despite its more complex structure, the
former also represents an essentially 1D chain of edge-sharing plaquettes.
Thus in \BICO, \LICO, and \CASTEL\ the Cu $2p$
main line is mostly due to local screening
processes, with negligible contribution
from the ZRS screening channel or
delocalization into O $2p$ band states and can therefore be assigned
to a \L\ final state using the language
of single-site models.

Comparing the spectra of  \S2CO\ and \SCO\ with those of the 
systems built of edge-sharing plaquettes, it is clear that the
former supports {\it no} locally screened final states
(i.e. peak $B$ is missing).
This conclusion is supported by recent
multiple-site calculations for \S2CO\
within the three-band $d$-$p$ model,\cite{okada96} in which no local
screening channel is predicted.
Based upon these calculations, feature $C^\prime$
represents a screening channel
specific to 1D systems in which one half
of the hole density pushed out from the core hole
site is transferred to the O sites above and
below the Cu core hole site, whereas the rest
is situated at the two extremes of the cluster chain due to
the large $t_{pd}$ hopping parameter.\cite{okada96}
The intensity of this 1D O $2p$ band
screening channel  depends on the charge transfer
\DEL\ between Cu and O.\cite{okada96}
Finally, we assign  feature
$A$ to the ZRS screened final state.
Although the calculation  reproduces the experimental
spectrum fairly well, the detailed shape of the main line of the
present high-resolution spectrum
and the position of $C^\prime$ in particular are not accurately
predicted.\cite{okada97}
This emphasizes that, even for this simplest chain
system, the frequently used $d$-$p$ model
based upon Cu $3d_{x^2-y^2}$ and O $2p_{x,y}$ orbitals is insufficient and 
probably has to be extended to include further Cu $3d$ states and the
inequivalency of the O sites in the chain.

The dependence upon the coupling path between the CuO$_4$ plaquettes 
also offers a simple explanation for the narrow Cu $2p$
main line observed in NaCuO$_2$.\cite{mizo} In this case, the plaquettes 
are edge-shared and in light of the data presented here
only a locally screened final state would be expected.

As regards the 2D systems, we note first that
the main line features $B$ have similar
positions to those in \LICO, \BICO, and \CASTEL\
but that additional features $A$ and $C$ appear.
Comparing the spectra of \BCOCL\ and \SCOCL\ and 
using arguments analogous to those presented above,
we attribute $A$ to the ZRS screened
final state and the intensity increase of $B$
compared to $A$ in \BCOCL\ to
the Cu$_B$ site present in the latter.
Since Cu$_B$ is connected
to the Cu$_A$O$_2$ network only via 90$^\circ$ Cu$_B$-O-Cu$_A$
coupling, the hopping matrix elements
are given by $t_{pp}$ and
therefore screening will mostly take place locally through the adjacent O
sites giving a \L\ final state. 
This interpretation is supported by the fact that
the difference between the scaled \BCOCL\ and \SCOCL\ Cu $2p$
spectra is very similar to those of \BICO, \LICO, and \CASTEL.
The introduction of an extra Cu site appears not to
affect the intensity of feature $C$, which we attribute
to a 2D O $2p$ band screening channel.\cite{okada95,myself}
For smaller square-planar clusters it has been
shown that the final states constituting
the O $2p$ band screening channel have little
overlap with the O sites neighboring the core hole
site and tend to spread over the whole cluster also
interacting only weakly with the Cu spins.\cite{okada95}
This justifies the notion of an O $2p$ band screening
in two dimensions distinct from the ZRS and local  screening
channels.
The different binding energies of $C$ could be related to the
different Cu-O bonding distances of $1.95$ \AA\
and $1.99$ \AA\ in \BCOCL\ and \SCOCL, respectively,
since the energy difference between the ZRS and the
O $2p$ band screened final states is expected to depend
among other parameters on the hopping energy $t_{pd}$.

Nevertheless, the large intensity of feature $B$ in \SCOCL\ (or in the
Cu$_A$O$_2$ part of the \BCOCL\ spectrum) 
remains wholly unexplained within the three-band
$d$-$p$ model in which the Cu $2p$ main line
should consist of only two features with large
spectral weight, which would approximately
reproduce features $A$ and $C$ in the experiment.\cite{myself}
An important lesson we derived from
the discussion of the much simpler linear chain
\S2CO\ was that the XPS final state is in fact
a highly excited state and it is probable that two-hole final
states of higher binding energies also have to be considered
involving for example
apical, i.e. non-planar, orbitals. A recent calculation
has shown that out-of-plane interactions can stabilize
higher binding energy two-hole states.\cite{raimondi}
The probability that these states
delocalize depends on the small hopping parameter $t_{pp}$ and
thus in accordance with the discussion above, they would lead to a 
locally screened final state.

In conclusion, we have presented high-resolution
Cu $2p$ XPS of 0, 1, and 2D high-quality single-crystalline
cuprates in which we distinguish between the 
locally and nonlocally screened contributions to the main lines.
In \BICO, \LICO, and \CASTEL, the main line is essentially due to
a locally screened final state.
Thus we predict that the lowest electron removal states
in these materials will be practically dispersionless in nature.  
In contrast, we could show that the chain systems
based upon corner-sharing CuO$_4$ plaquettes
(\SCO, \S2CO) display spectral weight due
only to nonlocal screening processes, which is in
qualitative agreement with calculations
within the three-band $d$-$p$ model.
The spectra of the two dimensional systems
(\SCOCL, \BCOCL) result from a combination of nonlocal
screening via ZRS formation and the
O $2p$ band, as well as a strong locally
screened contribution, in contradiction to calculations.
These data represent an ideal experimental
basis for the testing of models describing
core hole screening in correlated systems.

We gratefully acknowledge discussions
with K. Okada and  S.-L. Drechsler
and thank S. Oswald for generously
lending us machine-time.
Part of this work is supported by the BMBF
(05-605 BDA).
The work in Japan is supported by Grants for Priority
Area and for COE research from the Ministry
of Education, Science, Sports, and Culture.
%
%

%
%
%
%
\begin{figure}
\caption{Sketch of the Cu-O networks in the
different cuprates examined.
Cu atoms: $\bullet$; O atoms: $\circ$.
a) the linear chain in \S2CO,
b) the zigzag chain in
\SCO,  c) the chain of edge-shared plaquettes in \LICO, d) the
Cu$_A$O$_2$ plane in \SCOCL\
and the Cu$_3$O$_4$ plane  in \BCOCL\ containing
an extra Cu$_B$ site ($\Box$) and e) the crenellated chains of
\CASTEL.
}
\label{network}
\end{figure}
\begin{figure}
\caption{
Cu $2p_\frac{3}{2}$ photoemission
spectra of the single-crystalline cuprates.
The 1D chain systems a) \SCO, b) \S2CO, c) \LICO, and d) \CASTEL;
e) 0D \BICO; planar f) \SCOCL and g) \BCOCL.
Also shown is
the difference spectrum g)$-$f).
}
\label{cu2p}
\end{figure}
%
%
%
 \begin{table}
 \caption{Features of the Cu $2p_\frac{3}{2}$
 photoemission spectra of Fig.~\protect\ref{cu2p} obtained with
 a Voigt function fit:  included are the
 binding energies of the different 
 main line features, 
 their full width at half maximum (in brackets), 
 and the  ratio of
 the satellite to the main line intensity \ISIM. }
 \label{tab}
 \begin{tabular}{lddddd}
  Compound & $A$ & $B$& $C^\prime$& $C$
   & \ISIM\\
 \tableline
 \BICO & - & 934.1 & - & - & 0.58\tablenotemark[1] \\
 & & (1.6) & &  & \\
 \tableline
 \S2CO & 933.0   & - & 934.8   & - & 0.37\\
& (1.6) & &  (2.0) & & \\
 \SCO\ & 932.9  & - & 934.8  & - & 0.40\\
& (1.4) &  &  (2.2) &  & \\
 \LICO\ & - & 934.0   & - & -  &  0.56\\
&  &   (1.6) &  &   &  \\
 \CASTEL\ & - & 933.8  & - & -& 0.62\\
&  &  (1.1) &  & & \\
     \tableline
 \BCOCL\ & 932.7 & 933.8   & - & 935.7   & 0.50\\
 &  (0.8) &  (1.9) &  &  (2.2) & \\
 \SCOCL\ & 932.8 & 933.9   & - & 936.1  & 0.52\\
 & (1.0) & (2.2) & & (1.8) & \\
 Difference  & - & 933.9  & - &  935.5 & 0.45\\
  & & (1.2) & &  (1.5) & \\
\end{tabular}
\tablenotetext[1]{Emission from the Bi $4s$ core level contributes
at about 940 eV, consequently this value is an upper estimate. }
\end{table}
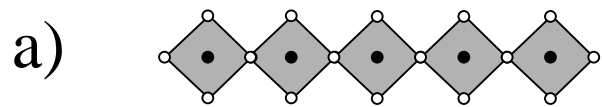
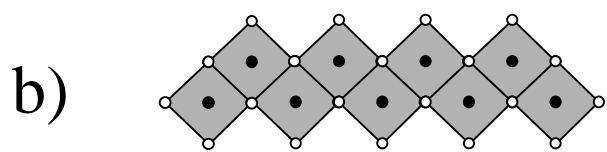
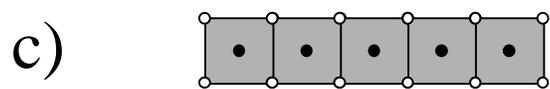
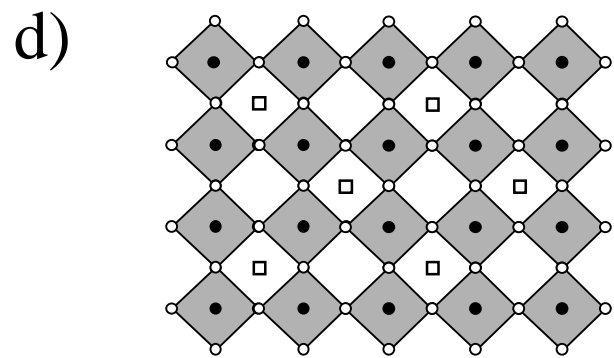
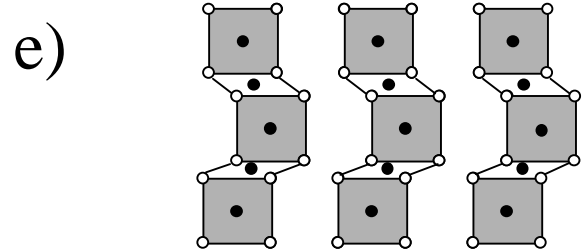
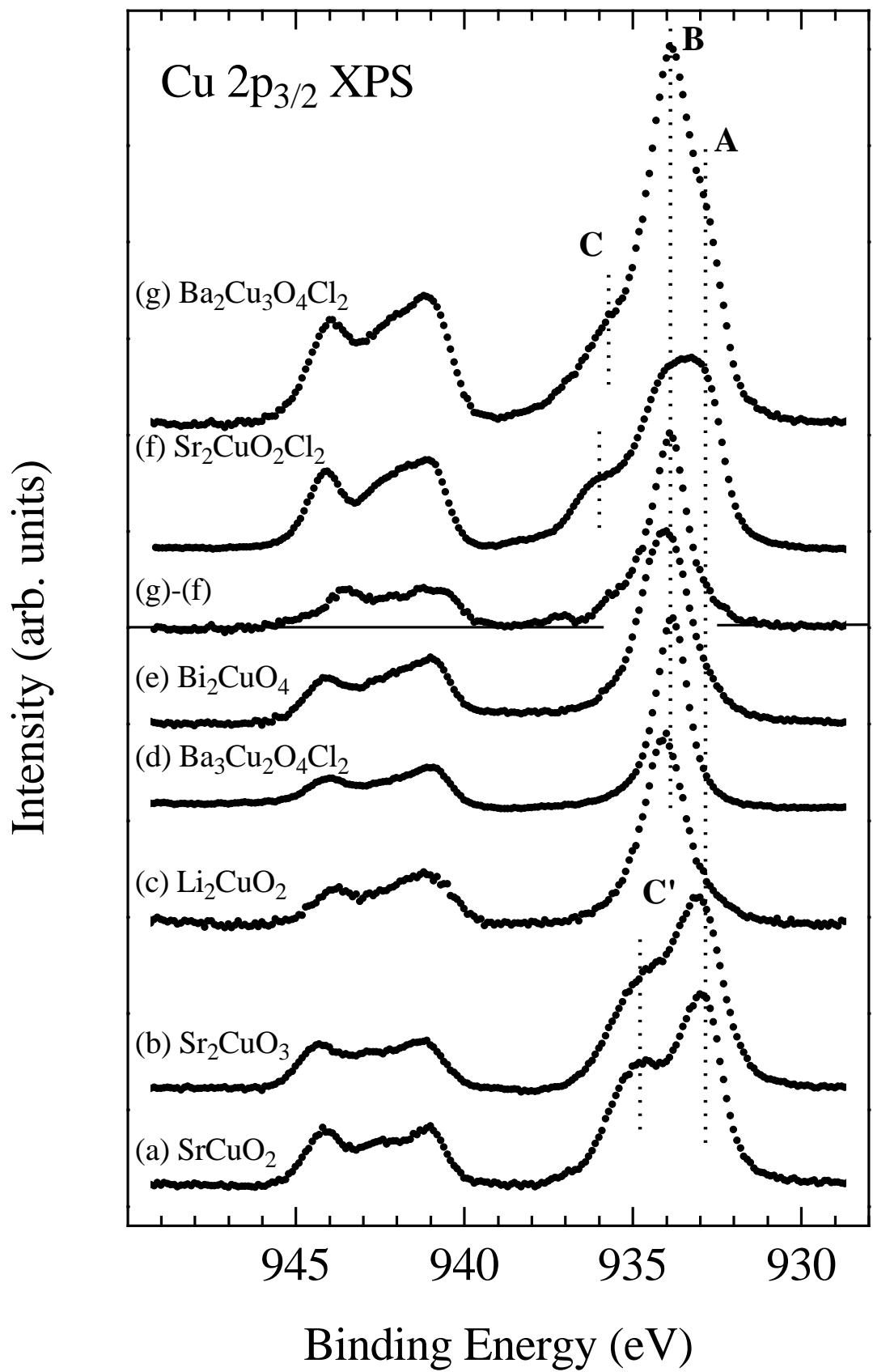
\end{document}